\documentclass[12pt,preprint]{aastex}
\usepackage[numberedappendix]{emulateapj5}  
 
\slugcomment{accepted for ApJ}

\shorttitle{Viscous dissipation in the ICM}
\shortauthors{Ruszkowski et al.}

\received{2003 October 27}
\begin{document}
 
\title{3D simulations of viscous dissipation in the intracluster medium}

\author{Mateusz Ruszkowski\altaffilmark{1}}
\affil{JILA, Campus Box 440,
University of Colorado at Boulder, CO 80309-0440; \email{mr@quixote.colorado.edu}}
                                                                                
\author{Marcus Br\"{u}ggen}
\affil{International University Bremen, Campus Ring 1, 28759 Bremen, Germany;
\email{m.brueggen@iu-bremen.de}}
\and
\author{Mitchell C. Begelman\altaffilmark{2}}
\affil{JILA, Campus Box 440, University of Colorado at Boulder, CO 80309-0440;
\email{mitch@jila.colorado.edu}}

\altaffiltext{1}{{\it Chandra} Fellow}
\altaffiltext{2}{also at Department of Astrophysical and Planetary Sciences,
University of Colorado at Boulder}


\begin{abstract}
We present three-dimensional simulations of viscous dissipation of
AGN induced gas motions and waves in clusters of galaxies. These
simulations are motivated by recent detections of ripples in the
Perseus and Virgo clusters. Although
the sound waves generated by buoyant bubbles decay with
distance from the cluster center, we show that these waves can contribute substantially
to offsetting the radiative cooling at distances
significantly exceeding the bubble size.  
The energy flux of the waves declines more steeply with radius
than the 
inverse-square law predicted by energy conservation, implying
that dissipation plays an important role in tapping the wave energy.
We show that such dispersing sound waves/weak shocks are
detectable as ripples on unsharp-masked X-ray cluster maps, and point out that 
the interfaces between the intracluster medium and old bubbles 
are also clearly detectable in unsharp-masked X-ray maps. This opens up 
the possibility of detecting 
fossil bubbles that are difficult to detect in radio emission.
This mode of heating is
consistent with other observational constraints,  such as the
presence of cool rims around the bubbles and the absence of strong
shocks. Thus,
the mechanism offers a way of heating clusters in a spatially
distributed and gentle fashion. We also discuss the energy transfer
between the central AGN and the surrounding medium. In our
numerical experiments, we find that roughly 65 per cent of the energy 
injected by the AGN is transferred to the intracluster medium and
approximately 25 percent of the injected  energy is dissipated
by viscous effects and contributes to heating of the gas. The
overall transfer of heat from the AGN to the gas is comparable to
the radiative cooling losses. The
simulations were performed with the FLASH adaptive mesh refinement
code.
\end{abstract}


\section{Introduction}
The long-standing problem of cooling flow clusters of
galaxies, in which the central cooling time is much shorter than
the Hubble time, is how to prevent the intracluster medium (ICM) from collapsing 
catastrophically on a short timescale. The original idea for maintaining the
overall cluster stability (Fabian 1994) was to postulate that a
certain amount of gas decouples from the flow and does not contribute
to the cooling of the  remaining gas. This model would require up
to 1000 $M_{\odot}$ yr$^{-1}$ in mass deposition rates to  guarantee
cluster stability. This has been found to be inconsistent with recent
{\it Chandra} (e.g., McNamara et al. 2000, Blanton et
al. 2001) and XMM-{\it Newton} observations (e.g., Peterson et
al. 2001, 2003; Tamura et al. 2001). {\it Chandra} observations
reveal a number of clusters with X-ray cavities/bubbles created by
the central active galactic nuclei (AGN). It has been suggested
by many authors that AGN feedback may play a crucial role in
self-regulating  cooling flows (e.g. Churazov et al. 2001,
Ruszkowski \& Begelman 2002, Brighenti \& Mathews 2003). One
of the main outstanding issues is how the AGN heating comes about in
detail. In principle, strong  shocks generated by AGN
outbursts can dissipate in the ICM and heat
the gas. However, imaging observations of cooling flow cores do
not give evidence for this mode of heating. Recent {\it Chandra}
observations of two well-known clusters,  the Perseus cluster
(Fabian et al. 2003a,b) and the Virgo cluster (Forman et
al. 2004), suggest that dissipation of sound waves and weak shocks
could be an important source of gas heating --- an idea first proposed
by Fabian et al. (2003a). Further support for the idea that
viscosity may play an important role in the ICM
comes from a recent study of density profiles in clusters (Hansen
\& Stadel 2003).  Recently, a number of papers have described
simulations of bubble-heated clusters (e.g., Churazov et al. 2001,
Br\"{u}ggen et al. 2002, Br\"{u}ggen \& Kaiser 2002, Br\"{u}ggen
2003, Quilis et al. 2001). Numerical simulations of viscous
dissipation of AGN energy in ICM were previously considered by
Ruszkowski et al. (2004) (Paper I) and Reynolds et al. (2004).\\
\indent The main purpose of this paper is to extend our
previous work on viscous heating of the ICM by waves to three
dimensions. The results of a simulation of viscous dissipation in
three dimensions could differ from our previous two-dimensional
results given that the amplitudes of waves decrease faster with radius in
three dimensions. This would directly affect the spatial
distribution of the viscous dissipation rate. 
Apart from performing our simulations in three dimensions,  we
extend our previous analysis and that of Reynolds et al. (2004) to
include aspects of heating that were previously neglected.  First,
we show that the conclusions drawn from our 2D simulations carry over to
three dimensions. Second, our new results include (i) X-ray maps and
unsharp-masked X-ray images, (ii) extend the discussion of spatial
distribution of energy dissipation and overall heating rate, (iii) present
details of the flow of heat between the bubbles and the intracluster
medium and (iv) discuss the wave decay rates.\\
\indent
The outline of
this paper is as follows. In the next section we describe the
assumptions of the model. Section 3 presents and discusses our
results, focusing on the  spatial distribution of heating. In
particular, we discuss the detectability of the ripples/sound
waves, their decay rate with distance from the center/bubble 
surface, the energy transfer from the AGN to the kinetic and thermal
energy  of the ICM, and the fraction of the dissipated energy that
heats the gas. The fourth section discusses the limitations of our model. 
The final section summarizes our findings.
\section{Assumptions of the Model}

The initial conditions, details of energy injection, dissipation and radiative 
cooling assumed in our simulations are similar to
those in Ruszkowski et al. (2003). Here we only summarize the 
differences from Paper I.\\
\indent
Calculations were done in three dimensions in Cartesian geometry using
the PPM adaptive mesh refinement code FLASH (version 2.3). 
Starting from a single
top-level block and using block sizes of 16$^3$ zones we allowed for 5
levels of refinement, giving an effective number of 256$^3$ zones. The
size of the computational domain was (200 kpc)$^3$, which corresponds
to an effective resolution of $\sim 0.8$ kpc. As previously, we
employed outflow boundary conditions on all boundaries.\\
\indent
When the source is active, each of the active regions
 has a constant luminosity of 
$L=1.6\times 10^{45}$ erg s$^{-1}$ and the rate at
which mass is injected into the active region is $\dot{\rho}V=2.8$
M$_{\odot}$yr$^{-1}$. The energy injection
is intermittent with a period of $3\times 10^7$ years, within which the
source is active for $0.5\times 10^{7}$ years.
Thus, since there are two active regions in the source, the time-averaged 
luminosity is $\sim 5.3\times 10^{44}$ erg s$^{-1}$.
The motivation behind a different choice 
of source parameters in the 3D case is as follows.
In 2D the code takes the input density (g cm$^{-3}$) and luminosity
density (erg cm$^{-3}$ s$^{-1}$) and interprets them as surface
quantities. That is,  the hydrodynamic equations are effectively integrated
along one axis by multiplying these quantities by a unit length.  As
the size of the active regions are smaller then unity, the code
effectively assigns a bigger volume to the active region when the
simulation is performed in 2D.  The source luminosity density that
enters the energy equation is a constant equal to $L/V_{\rm b}$, where $L$
is the source luminosity and $V_{\rm b}=(4/3)\pi r^{3}_{\rm b}$ is the
bubble volume (the same formula is used in the code in 2D and 3D
simulations). This means that in order to supply the same energy
to the cluster in 3D as in a 2D simulation, one has to convert 
the 2D luminosity to the 3D (real) luminosity using an approximate prescription
$L_{{\rm 3d}} \sim (3/4)$(unit length/$r_{\rm b}$)$L_{\rm{2d}}$.
The size of the active region in the 3D case was $r_{\rm b}=3$ kpc.
As far as the mass injection is concerned, we used higher values for
purely pragmatic reasons. Higher injection rates assure that the
densities in the active regions are reduced to a lesser degree then
they would have been if the injection rate had been lower. This means that 
the constraints on the Courant timestep are less severe and the simulations 
can be performed in a shorter time.\\
\indent
We use the standard Spitzer viscosity for an
unmagnetized plasma  (Braginskii 1958), for which $\mu =
7.1\times 10^{-17}(\ln\Lambda/31)^{-1}T^{5/2}$ g cm$^{-1}$ s$^{-1}$.
As conditions inside the buoyantly rising bubbles
are very uncertain and because we want to focus on energy
dissipation in the ambient ICM, we assume that dissipation occurs only
in the regions surrounding the buoyant gas. To this end we impose
a condition that switches on viscous effects provided that the
fraction of the injected gas in a given cell is smaller than
$10^{-3}$. The gas obeys a 
polytropic equation of state with an adiabatic index of $\gamma=5/3$.
Unlike in Paper I we do not assume that the active regions are initially
underdense and overheated. 

\section{Results}

\subsection{Morphology of X-ray Emission and Dissipation}
The top row in Figure 1 shows X-ray maps of the heated cluster. The panels show five
different epochs from earliest on the left to the latest on the
right. We assumed that the main contributor to  X-ray emissivity
is free-free emission. The maps correspond to emission integrated in
the energy band $E\in (2-10)$ keV. The axis of injection is
located in the plane of the sky. The emissivity contrast between
the injected material and the surrounding  medium is strongest at
the center. As the bubbles rise buoyantly in the cluster
atmosphere, the contrast diminishes and so does the probability of
detecting the bubbles. \\
\indent
The middle row in Figure 1 presents unsharp-masked X-ray images
corresponding to the images on the top row. These images were
generated by smoothing the original X-ray map and subtracting the
original X-ray map. Smoothing was done by convolving the original
image with a Gaussian filter centered on a given point. The full
width at half maximum of the adopted Gaussian distribution was
$\sim 6$ kpc. It is
evident from this figure that ripples outside bubble locations are
visible. As discussed below, these perturbations are
the sound waves/weak shocks. Note also that the interfaces between the bubble location
and the ICM are clearly present in these maps. This means that the ripples 
in the Perseus cluster can be due to a combination of sound waves, weak shocks and interfaces between 
the ICM and fossil bubbles.\\ 
\indent
The bottom row in Figure 1 shows the viscous dissipation rate
(erg s$^{-1}$g$^{-1}$) associated with the dispersing waves seen in
the middle row. These maps show cross-sections through the cluster center
that are perpendicular to the line of sight.

\subsection{Spatial Distribution of Heating and Total Dissipation}
In Figure 2 we show the ratio of the viscous heating rate
to the radiative cooling rate as a function of time for a range
of radii. More distant regions are heated at later times and hence the
curves corresponding to these regions rise at later times. 
It is interesting to note that the heating-to-cooling
ratio is of order unity even for more distant regions close to the
cooling radius. Thus, heating is well distributed spatially, even
though the bubbles occupy a smaller volume than the waves. Note that 
the curves exhibit a clear periodic behavior which is due to the
intermitency of the central source. We also note that this figure is
similar to Fig. 2 in Paper I, thus confirming that 
one of the main conclusions of their paper holds in three-dimensional 
geometry.\\
\indent
Figure 3 shows the ratio of viscous heating to radiative cooling rate as a
function of distance from the cluster
center for equally-spaced time intervals of
$\Delta t=10^{7}$ yr until $2\times 10^{8}$ yr.
As time increases the curves start to decline at progressively larger radii.
From this figure one can also deduce the characteristic speed of
the wave pattern (see below) and compare it with analytical estimates.\\
\indent
Figure 4 shows the ratio of the volume-integrated heating
rate to the volume-integrated cooling rate as a function of
time. Note that this ratio is of order unity. Therefore,
this heating mechanism has the potential for significantly
affecting the rate at which the gas loses its internal energy or
perhaps even offsetting radiative cooling altogether.\\
\indent
We stress that the actual value of the ratio of heating to cooling (Figs. 2--4) 
depends on the parameters
adopted in the model. However, fine tuning may not be necessary 
as already argued by Paper I.\\
\indent
Additional heating may result from damping of large scale motions caused
by entrainment and lifting of the gas surrounding the bubbles. This is
the general mechanism  proposed by Begelman (2001) and Ruszkowski
\& Begelman (2002) in their discussion of ``effervescent
heating''. The widespread spatial distribution of heat in
this mechanism can be achieved when the buoyant bubbles rise in
the ICM perturbed by preceding bubbles. That is, subsequent
bubbles find lower resistance to move in all directions once they
encounter underdense regions ``drilled'' by earlier bubbles.
Moreover, when the bubbles move a substantial distance from the
cluster center, the lateral spreading is further enhanced as 
the bubble entropy becomes comparable to the entropy of the
ambient medium. Spatial spreading
of heat could also be facilitated if the jet (or other form of outflows) precesses 
or if the black hole
in the center of the AGN adjusts in response to a change in the orientation of the accretion
disk. This could be triggered by a merger with a substructure clump
in the cluster or by the oscillatory motion of the AGN around
the cluster center of gravity (see e.g. Johnstone et al. 2002).

\subsection{Energy Transfer Between the Bubbles and the ICM}
In Figure 5 we present the cumulative
injected energy $E_{\rm inj}$ (solid curve), energy contained in the
rising bubbles $E_{\rm bubb}$ (dashed) and the energy transferred
to the ICM ($E_{\rm tran}\equiv E_{\rm inj}-E_{\rm bubb}$; dotted). The bubble energy 
$E_{\rm bubb}$ is defined as

\begin{eqnarray}
E_{\rm bubb} & \equiv & \int_{\rm bubb}
  [e_{\rm pot}(t)+e_{\rm int}(t)+e_{\rm kin}(t)]\rho dV\nonumber\\
             & -      & [e_{\rm pot}(0)+e_{\rm int}(0)]\rho_{0} (2V_{0}) ,
\end{eqnarray}

\noindent
where $e_{\rm pot}$, $e_{\rm kin}$ and $e_{\rm int}$ are the gravitational potential,
specific kinetic energy density and specific internal energy density, respectively.
The $2V_{0}$ factor is the total volume of the initial injection regions.
The integration is performed over the volume of the bubbles. 
The bubble energy increases during
outbursts and then decreases during dormant phases. 
The decrease of the energy in the bubbles with time can be attributed to
two factors: (i) energy is transferred to the ambient medium
via $PdV$ work and because the rising bubble experiences drag from the
surrounding gas and (ii) mixing of the
bubble and the ICM gas. Recall that the bubble is defined as the region
where the fraction of the injected gas is greater than $10^{-3}$ 
(this region overlaps exactly with the low density bubble region).
This assures that essentially no energy injected into the bubble is omitted in
the calculation of the bubble energy. The contribution to the bubble energy
from the gravitational potential is relatively small. Since mixing occurs without
a significant change in pressure and the bubble internal energy is
proportional to pressure, the change in bubble energy comes mostly
from the $PdV$ work done by the bubble against its surroundings
and the drag on the bubble from the ambient ICM.\\
\indent
The fraction of the AGN energy that is transferred to the
surrounding ICM is plotted in Figure 6 (solid curve). The ratio
of viscously dissipated energy to the energy
injected by the AGN is plotted as the bottom (dashed) curve in Figure
6. This figure indicates that approximately 35 per cent of the
energy injected by the AGN remains in the buoyant bubbles while the
rest is transferred to the surrounding ICM and that about 25 per cent of the
injected energy ends up heating the ambient ICM viscously.
We point out that not all of the energy that is transferred to the ICM
is converted to heat. It is viscous dissipation that is responsible for
heating the gas.
The fraction of the injected energy that is transferred to the ambient
ICM agrees qualitatively with simple analytical estimates presented in
Paper I.
For the adopted adiabatic index and assuming pressure equilibrium between the bubble
and the ICM about $40\%$ of the input energy can be transferred to the ambient
gas. 
Because the cavities are mildly overpressured, the fraction of the
input power transferred to the ICM in the actual simulation is a
little larger.

\subsection{Wave decay}
From Figure  3 one can deduce the characteristic speed of
the wave pattern which we found to have a Mach number of $\sim 1.3$. 
Thus, the waves can be interpreted as strong sound waves or weak shocks.
In the weakly nonlinear regime, the speed of the wave is 
$c_{s}(1+[(\gamma +1)/2]\alpha)$, where $c_{s}$ is the sound speed and 
$\alpha\sim\delta\rho/\rho$ is the normalized wave density amplitude 
(Stein \& Schwartz 1972, Mihalas \& Mihalas 1984).
For the typical amplitudes seen in the simulation, the wave speed of 
Mach 1.3 is consistent with these estimates.\\
\indent
In Figure 7 we show the energy flux of the decaying wave
corresponding to the initial outburst. The simulation results are
denoted by filled squares connected by a solid line. Also shown
for reference is the decay profile corresponding to $\sim r^{-2}$. All
curves have arbitrary units. 
The period-averaged wave energy per unit time that is
streaming through a surface $S$ in a direction perpendicular to
this surface is $L_{w}\sim (\delta P)^{2}S/(\rho v_{w})$, where
$v_{w}$ is the wave speed. In the absence of any dissipation, the
energy flux should scale as $\sim r^{-2}$.
However, the slope of the energy flux in our simulations
is steeper. This means that viscous dissipation plays an important role in
tapping the wave energy. \\
\indent
The characteristic dissipation length $l$ can be estimated from
$l\sim 70\lambda_{10}^{2}n_{0.02} T_{4}^{-2}$ kpc, where
$\lambda=10\lambda_{10}$kpc, $n=0.02n_{0.02}$ cm$^{-3}$ 
and $T=4T_{4}$ keV (Fabian et al. 2003a, cf. Landau and Lifshitz 1975). 
The dissipation length can also be estimated from 
$l\sim\ [\partial\ln(\rho\alpha^{2})/\partial r + 2/r]^{-1}$,
where $\alpha\equiv \delta\rho/\rho$ is the normalized density amplitude of the wave
and can be directly derived from the simulation results.
At $r\sim 55$ kpc, the dissipation length is of order 40 kpc, which is
qualitatively consistent with the above simple analytical estimates.

\subsection{Caveats}
We stress that our
use of the Spitzer viscosity is meant to be illustrative and may
not accurately represent the momentum transport in the magnetized
intracluster medium.  For one thing, magnetic shear stresses are
likely to dominate over molecular viscosity in the transport of
bulk momentum.  This could either enhance or suppress the
dissipation of sound waves, and will almost certainly make the
dependence of stress on the velocity field more complicated. For
another, in this macroscopic form of momentum transport the rate
of dissipation (due to reconnection) would be nonlocally related to
the stress tensor.  Treatment of these effects will require
high-resolution magnetohydrodynamical simulations.  Moreover,
magnetic fields could introduce effects similar to bulk viscosity,
as a result of plasma microinstabilities.  In our simulations we
neglected bulk viscosity since it vanishes for an ideal gas. We
note that bulk viscosity, if present, could dissipate waves even
more efficiently. Finally, we have neglected the effects of
thermal conduction, which (assuming Spitzer conductivity) could
damp the sound waves more quickly than Spitzer viscosity (since
the conductive dissipation rate exceeds the viscous one by a factor
$\sim 10$ under the simplified assumption that waves are
linear and that the gas has constant density and pressure and
gravity can be neglected; see Landau and Lifshitz 1975). We
point out that, as long as the waves are linear, the nature of the
wave decay due to Spitzer viscosity or Spitzer conductivity is the
same, i.e., the only change is the constant damping coefficient. 
Note that the waves considered here are either linear (sound
waves) or weakly non-linear (weak shocks) and characterized
by a relatively small Mach number. Since conductivity is expected
to be suppressed by magnetic fields, a realistic assessment of
whether conduction enhances the damping rate of sound waves is
beyond the scope of this investigation.
If the dissipation rate significantly exceeds the Spitzer
value, then the sound waves will be damped more efficiently and only the
gas close to the bubbles will be heated efficiently.\\
\indent
We now argue that the choice of boundary conditions has negligible 
effects on our results.
First of all, the issue of reflection has no effect at all on our
results presented in Figures 4, 5, 6 and 7. This is because the waves
had no time to reach the boundary for the times considered in these
figures.\\
\indent
The waves reach the boundary approximately at time $\sim 6\times 10^{7}$ yr. However, 
there is
no clear rapid rise in the dissipation rate for the times before and
after this time. No such jump is seen in Figure 2 (e.g., for the
curves that correspond to outer shells in the cluster, where the
reflection effects should be seen immediately) or in Figure 4, which
shows the total dissipation rate. Had the reflection been important,
such a jump would have been clearly visible.\\
\indent
In the last row of Figure 1, 2nd (earlier time) and 3rd (later time) columns, shows the
dissipation wave in the process of crossing the boundary. No
reflection is seen in either 2nd or 3rd column figure. Also,  the wave
fronts in the corners propagate across the boundary without any
obvious signs of strong reflection.\\
\indent
It is known that 
for supersonic fluctuations the outflow boundary conditions are 
exact and no reflections
are expected.  The fact that our waves are strong sound waves or weak
shocks traveling at mildly supersonic velocities helps to minimize
reflections.\\
\indent
In principle, the outflow boundary conditions can be made arbitrarily
close to the exact ones if the simulation resolution is sufficiently
high. That is, the error made by replacing the ghost zone solution
based on some exact method by the value copied from the last active
zone (outflow boundary condition) tends to zero as the resolution
increases. Similarly, when the perturbations in question are smoother,
the outflow boundary conditions give a more accurate answer because
the waves are better resolved. We note that the effect of dissipation
is to disperse the waves and make them more easily
resolvable. Moreover, the amplitude of the waves decays faster in 3D
than in 2D. Thus, in 3D the effects of reflection are reduced.\\
\indent
The outflow boundary conditions are exact for a wave traveling along
the boundary. Therefore, the degree to which a wave is reflected
decreases for waves moving at an angle to the boundary.

\section{Summary}
To summarize, we have analyzed the energy deposition in the cluster due to
rising bubbles, sound waves and weak shocks. 
This was motivated by the recent discovery of such waves in the Perseus
cluster by Fabian et al. (2003a) and in the Virgo cluster by Forman et al. (2004).
We found that the dissipated energy may be comparable to the cooling rate,
thereby significantly affecting the cooling flow or even quenching it
altogether. We showed that about 65 per cent of the energy injected by the central
source can be transferred to the ICM.
Approximately 25 per cent of the energy injected by the AGN
can be converted to heat, assuming Spitzer viscosity. 
We discussed the wave decay rates and showed that a significant fraction of
wave energy is deposited within the cooling radius. The computed decay
rates are consistent with linear theory estimates of the damping
length. The damped sound waves or weak shocks are still detectable
in unsharp-masked X-ray images. 
Old bubbles become increasingly difficult to detect in the
X-ray maps as the contrast between the rising bubbles and the surrounding
gas diminishes.
However, apart from sound waves and weak shocks in 
unsharp-masked X-ray maps,
the interfaces between the intracluster medium and old bubbles 
are also clearly visible. This opens up the possibility of detecting 
fossil bubbles that are difficult to detect in radio emission.


\acknowledgments{
MR thanks Daniel Proga for discussions.
Support for this work was provided by National 
Science Foundation grant AST-0307502 and the National Aeronautics and Space
Administration through {\it Chandra} Fellowship Award Number PF3-40029
issued by the Chandra X-ray Observatory Center, which is operated by
the Smithsonian Astrophysical Observatory for and on behalf of the
National Aeronautics and Space Administration under contract
NAS8-39073.
MB acknowledges support by DFG contract BR2026/2 and thanks
JILA, University of Colorado at Boulder for their hospitality. 
The software used in this work was in part developed by the DOE-supported
ASCI/Alliance Center for Astrophysical Thermonuclear Flashes at the
University of Chicago. Preliminary computations were performed
on JILA Keck cluster sponsored by the Keck Foundation.}

\newpage

\begin{figure}
\centerline{
\includegraphics[width=7.0in,angle=0]{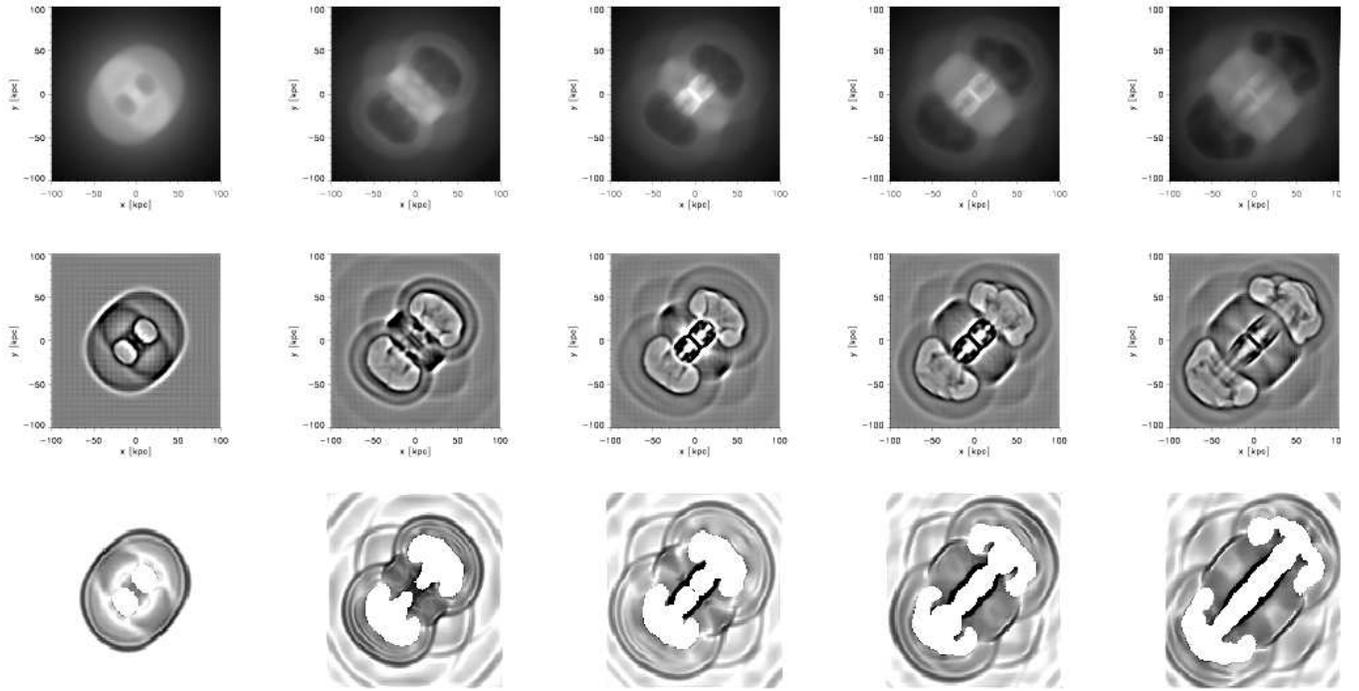}}
\caption{\footnotesize 
Top row shows the X-ray emissivity maps of the AGN-heated cluster.
Snapshots correspond to $3.0\times 10^{7}$, $1.15\times 10^{8}$,
$1.25\times 10^{8}$, $1.55\times 10^{8}$ and $1.85\times 10^{8}$ years, 
from left to right, respectively.
Middle row shows X-ray unsharp masked maps corresponding to  X-ray maps.
Bottom row shows the map of the viscous dissipation pattern.
Whereas the maps in the bottom row show cross-sections through the cluster center
that are perpendicular to the line of sight, the first two rows correspond to 
projections onto the plane of the sky.
\label{fig1}} 
\end{figure}

\begin{figure}
\centerline{
\includegraphics[width=3.5in,angle=90]{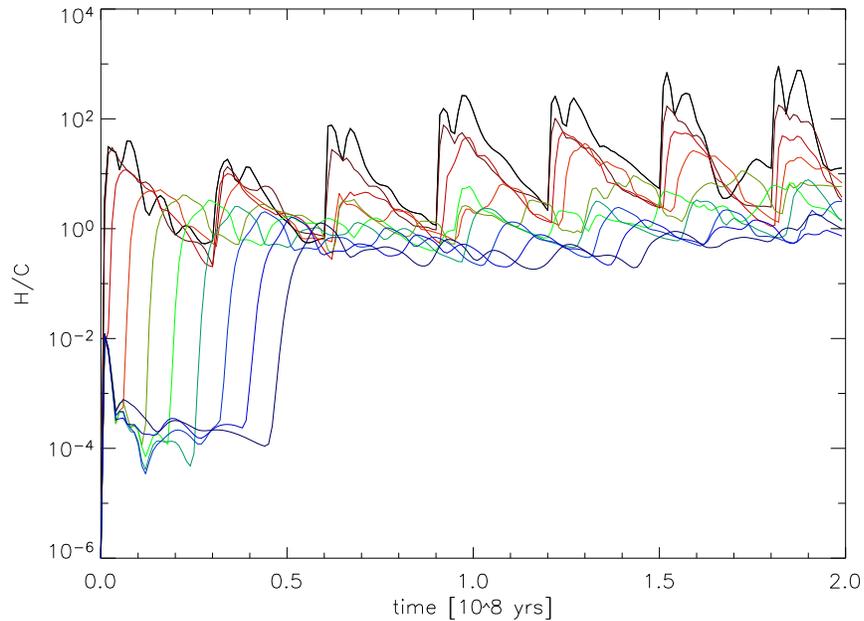}}
\caption{\footnotesize The ratio of viscous heating to radiative cooling rate as a
function of time for a number of concentric shells around the cluster
center.  The curves that start rising at later times correspond to
shells located further away from the center. The heating-to-cooling
ratio was calculated in ten shells starting from the first shell at 5
kpc and the remaining shells located in increments of 10 kpc away from
the cluster center. Note that the heating rate is comparable to the
cooling rate.
\label{fig2}}
\end{figure}

\begin{figure}
\centerline{
\includegraphics[width=3.5in,angle=90]{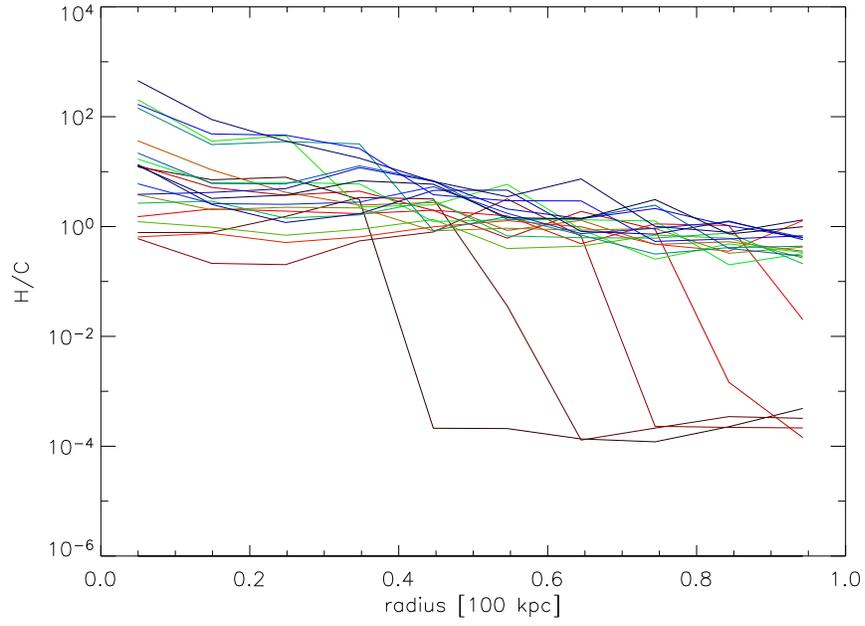}}
\caption{\footnotesize 
The ratio of viscous heating to radiative cooling rate as a
function of distance from the cluster
center for equally-spaced time intervals of $\Delta t=10^{7}$ yr until $2\times 10^{8}$ yr.
\label{fig3}}
\end{figure}

\begin{figure}
\centerline{
\includegraphics[width=3.5in,angle=90]{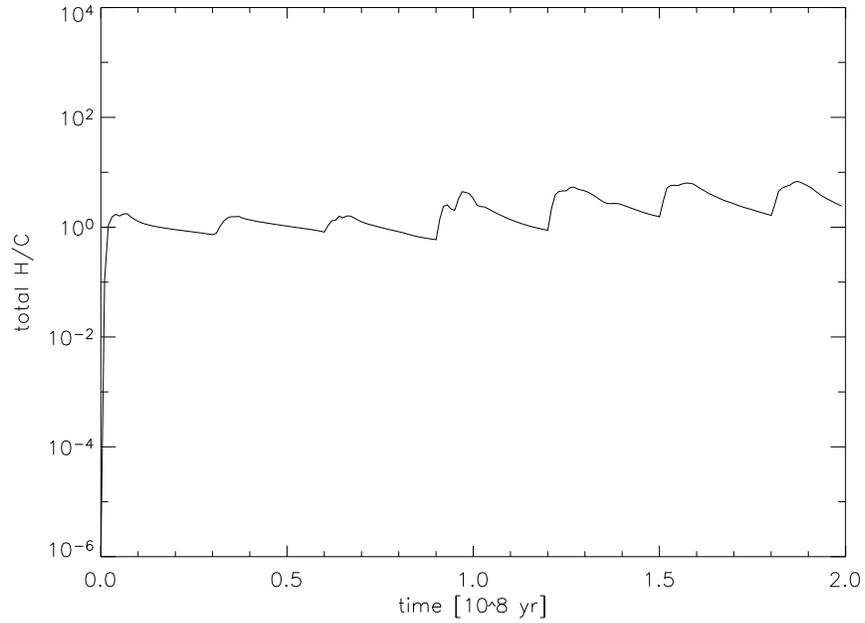}}
\caption{\footnotesize The ratio of volume-integrated heating rate (within
100 kpc from the center) to volume-integrated cooling rate
as a function of time.
\label{fig4}}
\end{figure}

\begin{figure}
\centerline{
\includegraphics[width=3.5in,angle=90]{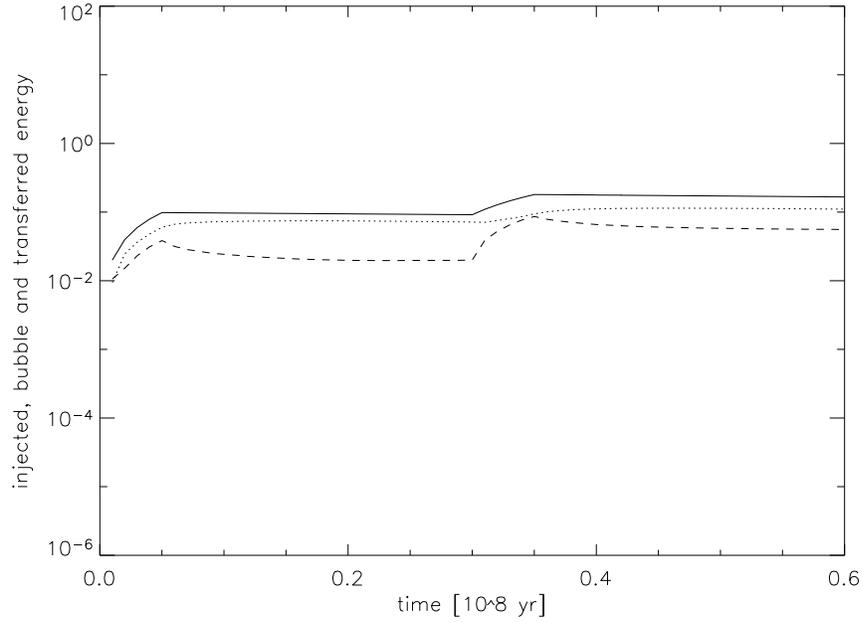}}
\caption{\footnotesize The cumulative injected energy $E_{\rm inj}$ (solid curve), 
bubble energy $E_{\rm bubb}$ (dashed curve) and 
energy transferred to the ambient ICM ($\equiv E_{\rm inj}-E_{\rm bubb}$; dotted curve) 
as a function of time. All plots are in arbitrary units. 
We only plot data until 60 Myr as for later times 
the waves start to escape the computational box.
\label{fig5}}
\end{figure}

\begin{figure}
\centerline{
\includegraphics[width=3.5in,angle=90]{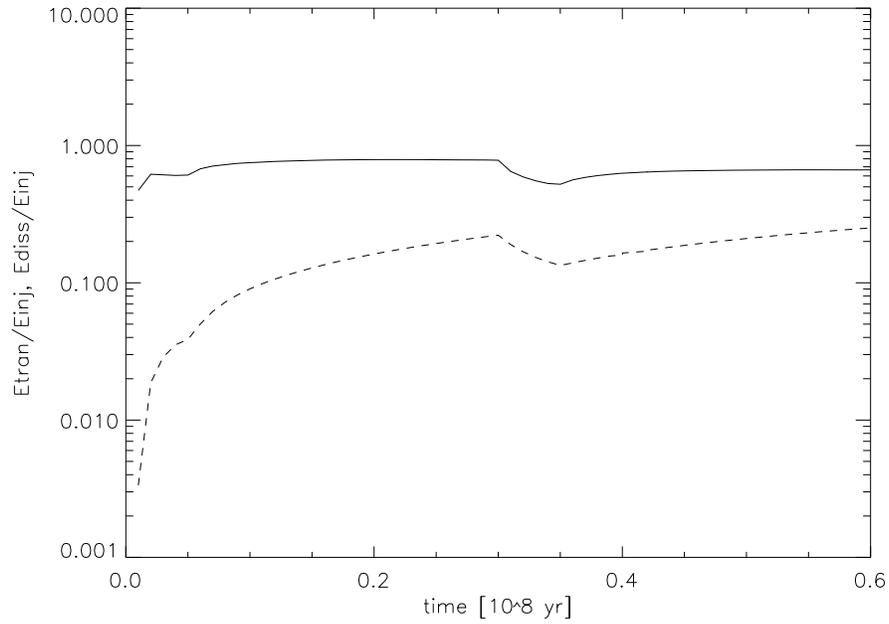}}
\caption{\footnotesize The ratio of cumulative transferred energy to the cumulative injected 
energy (top curve)
and the ratio of the cumulative viscously dissipated energy to the cumulative injected energy
as a function of time. 
We only plot data until 60 Myr as for later times 
the waves start to escape the computational box.
\label{fig6}}
\end{figure}

\begin{figure}
\centerline{
\includegraphics[width=3.0in,angle=270]{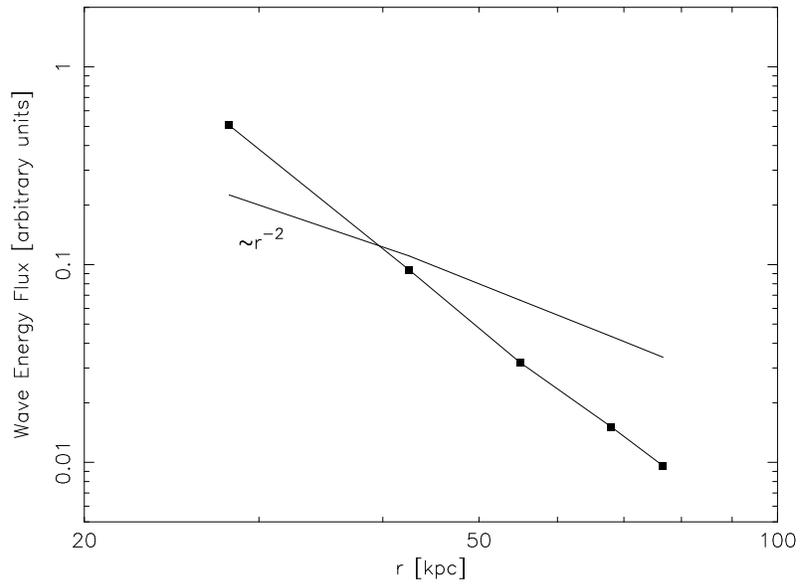}}
\caption{\footnotesize The wave energy flux as a function of the distance from the cluster center.
The $\sim r^{-2}$  
decay profile is shown for comparison. All quantities are in arbitrary units.
\label{fig7}}
\end{figure}

\end{document}